%% file: ms.tex
\documentclass[10pt,a4paper,twoside]{article}
\usepackage[a4paper,left=30mm,right=30mm, top=1cm, bottom=2cm]{geometry}

\pdfoutput=1
\usepackage[T1]{fontenc}
\usepackage[utf8]{inputenc} 
\usepackage[english]{babel} 
\usepackage{graphicx}
\usepackage[nottoc,numbib]{tocbibind}
\usepackage{hyperref}
\usepackage{pdfpages} 
\usepackage{caption}
\usepackage{authblk}
\usepackage{verbatim}

\date{\today}

\begin{document}

\input{01-title_abstract}


\input{10-introduction}

\input{30-supply_chain}
\input{40-vulnerabilities}

\input{50-mitigation}

\input{60-conclusion}

\input{70-literature}

\end{document}

%% file: 01-title_abstract.tex
\title {Vulnerabilities of Connectionist AI Applications: Evaluation and
Defence}
\author[1,2,*]{Christian Berghoff}
\author[1]{Matthias Neu}
\author[1,2,*]{Arndt von Twickel}
\affil[1]{Federal Office for Information Security, Bonn, Germany}
\affil[2]{Corresponding authors: \{christian.berghoff, arndt.twickel\}@bsi.bund.de}
\affil[*]{These authors contributed equally}
	\maketitle 
\begin{abstract}
This article deals with the IT security of connectionist artificial intelligence (AI) applications, focusing on threats to integrity, one of the three IT security goals. Such threats are for instance most relevant in prominent AI computer vision applications. In order to present a holistic view on the IT security goal integrity, many additional aspects such as interpretability, robustness and documentation are taken into account. A comprehensive list of threats and possible mitigations is presented by reviewing the state-of-the-art literature. AI-specific vulnerabilities such as adversarial attacks and poisoning attacks as well as their AI-specific root causes are discussed in detail. Additionally and in contrast to former reviews, the whole AI supply chain is analysed with respect to vulnerabilities, including the planning, data acquisition, training, evaluation and operation phases. The discussion of mitigations is likewise not restricted to the level of the AI system itself but rather advocates viewing AI systems in the context of their supply chains and their embeddings in larger IT infrastructures and hardware devices. Based on this and the observation that adaptive attackers may circumvent any single published AI-specific defence to date, the article concludes that single protective measures are not sufficient but rather multiple measures on different levels have to be combined to achieve a minimum level of IT security for AI applications.\\
\end{abstract}



%% file: 10-introduction.tex
\section{Introduction}\label{sec:introduction}
This article is concerned with the IT security aspects of artificial intelligence (AI) applications\footnote{AI is here defined as the capability of a machine to either autonomously take decisions or to support humans in making decisions. In order to distinguish AI from trivial functions, such as, for instance, a sensor that directly triggers an action using a threshold function, one might narrow the definition to non-trivial functions but since this term is not clearly defined, we refrain from doing so.}, namely their vulnerabilities and possible defences. As any IT component, AI systems may not work as intended or may be targeted by attackers. Care must hence be taken to guarantee an appropriately high level of safety and security. This applies in particular whenever AI systems are used in applications where certain failures may have far-reaching and potentially disastrous impacts including the death of people. Examples commonly cited include computer vision tasks from biometric identification and authentication as well as driving on-road vehicles at higher levels of autonomy \cite{sae2018}. Since the core problem of guaranteeing a secure and safe operation of AI systems lies at the intersection of the areas of AI and IT security, this article targets readers from both communities. 

\subsection{Symbolic vs. connectionist AI}
AI systems are traditionally divided into two categories: symbolic AI (sAI) and non-symbolic (or connectionist) AI (cAI) systems. sAI has been a subject of research for many decades, starting from the 1960s \cite{Lederberg1987}. In sAI, problems are directly encoded in a human-readable model and the resulting sAI system is expected to take decisions based on this model. Examples of sAI include rule-based systems using decision trees (expert systems), planning systems and constraint solvers. In contrast, cAI systems consist of massively parallel interconnected systems of simple processing elements, similar in spirit to biological brains. cAI includes all variants of neural networks such as deep neural networks (DNNs), convolutional neural networks (CNNs) and radial basis function networks (RBFNs) as well as support-vector machines (SVMs). Operational cAI models are created indirectly using training data and machine learning and are usually not human-readable. The basic ideas for cAI systems date back to as early as 1943 \cite{McCulloch1943}. After a prolonged stagnation in the 1970s, cAI systems slowly started to gain traction again in the 1980s \cite{Haykin1999}. In recent years, starting from about 2009, due to significant improvements in processing power and the amount of example data available, the performance of cAI systems has tremendously improved. In many areas, cAI systems nowadays outperform sAI systems and even humans. For this reason, they are used in many applications, and new proposals for using them seem to be made on a daily basis. Besides pure cAI and sAI systems, hybrid systems exist. In this article, sAI is considered a traditional IT system and the focus is on cAI systems, especially due to their qualitatively new vulnerabilities that in turn require qualitatively new evaluation and defence methods. Unless otherwise noted, the terms AI and cAI will from now on be used interchangeably.

\subsection{Life cycle of AI systems}
\begin{figure}
	\centering
	\includegraphics[width=\textwidth]{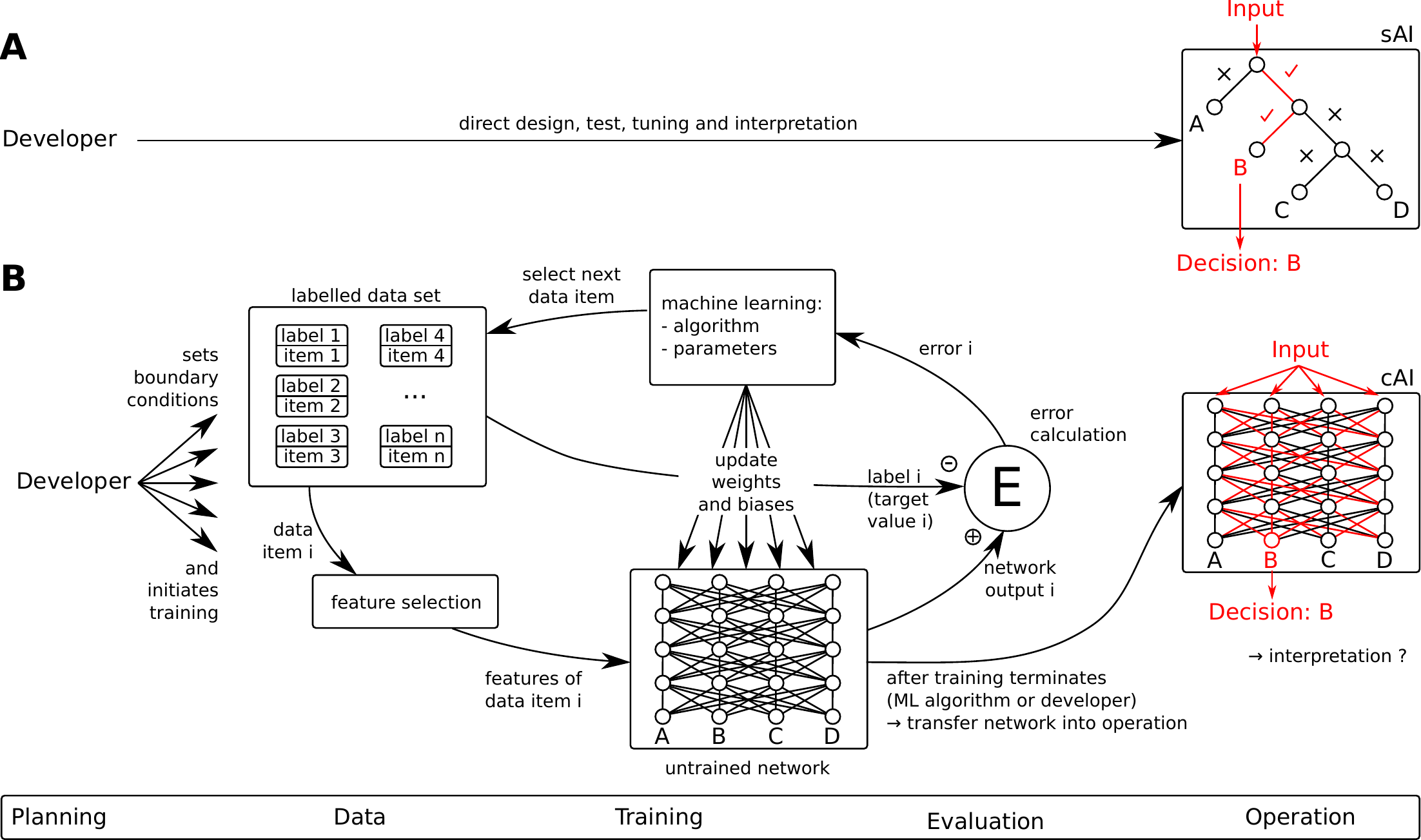}
	\caption{Contrasting the development of (A) symbolic AI (sAI) and (B) connectionist AI (cAI) systems. Whereas sAI systems are directly designed by a human developer and are straightforward to interpret, cAI systems are trained by means of machine learning (ML) algorithms using large data sets (this figure shows supervised learning using a labelled data set). Due to their indirect design and their distributed decision-making, cAI systems are very hard to interpret.}
	\label{img:Figure01}
\end{figure}
In contrast to sAI and traditional IT systems, cAI systems are not directly constructed by a human programmer (cf. \autoref{img:Figure01}). Instead, a developer determines the necessary boundary conditions, i.e. required performance\footnote{In contrast to narrowing the term performance to cover only accuracy, we use it in a broader sense, cf. \autoref{sec:planning} for details.}, an untrained AI system, training data and a machine learning (ML) algorithm, and then starts a ML session, during which a ML algorithm trains the untrained AI system using the training data. This ML session consists of alternating training and validation phases (not shown in \autoref{img:Figure01}) and is repeated until the required performance of the AI system is achieved. If the desired performance is not reached within a predefined number of iterations or if performance ceases to increase beforehand, the training session is cancelled and a new one is started. Depending on the ML policy, the training session is initialised anew using randomised starting conditions or the boundary conditions are manually adjusted by the developer. Once the desired performance is achieved, it is validated using the test data set, which must be independent from the training data set. Training can be performed in the setting of supervised learning, where the input data contain preassigned labels, which specify the correct corresponding output (as shown in \autoref{img:Figure01}), or unsupervised learning, where no labels are given and the AI system learns some representation of the data, for instance by clustering similar data points. While this article takes the perspective of supervised learning, most of its results also apply to the setting of unsupervised learning. After successful training, the AI system can be used on new, i.e. previously unknown, input data to make predictions, which is called inference.

Due to this development process, cAI systems may often involve a complex supply chain of data, pre-trained systems and ML frameworks, all of which potentially impact security and, therefore, also safety. It is well known that cAI systems exhibit vulnerabilities which are different in quality from those affecting classical software. One prominent instance are so-called adversarial examples, i.e. input data which are specially crafted for fooling the AI system (cf. \autoref{sec:operation}). This new vulnerability is aggravated by the fact that cAI systems are in most practical cases inherently difficult to interpret and evaluate (cf. \autoref{sec:black_box}). Even if the system resulting from the training process yields good performance, it is usually not possible for a human to understand the reasons for the predictions the system provides. In combination with the complex supply chain as presented in \autoref{sec:supply} this is highly problematic, since it implies that it is not possible to be entirely sure about the correct operation of the AI system even under normal circumstances, let alone in the presence of attacks. This is in analogy to human perception, memory and decision-making, which are error-prone, may be manipulated (\cite{Eagleman2001,Loftus2005,Wood2013}, cf. also \autoref{img:Figure06}) and are often hard to predict by other humans \cite{Sun2018}. As with human decision-making, a formal verification of cAI systems is at least extremely difficult, and user adoption of cAI systems may be hampered by a lack of trust.

\subsection{IT security perspective on AI systems}
\begin{figure}
	\centering
	\includegraphics[width=0.8\textwidth]{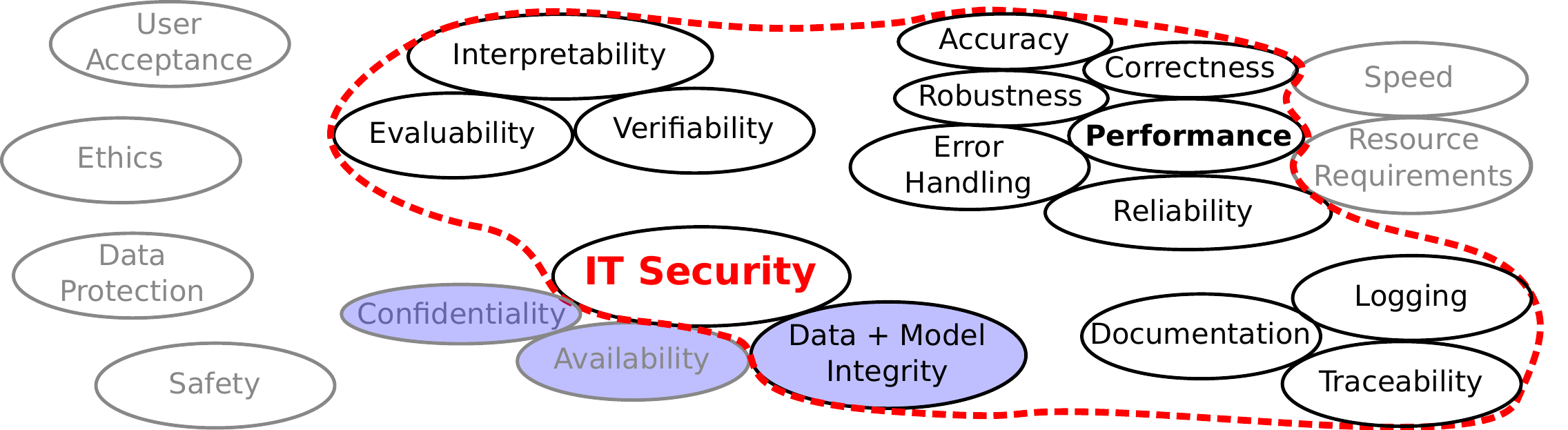}
	\caption{Besides the three core properties confidentiality, integrity and availability, a holistic view on the IT security of AI applications involves many additional aspects. This paper focuses on data and model integrity and important related aspects, especially robustness, interpretability and documentation, here depicted in the centre and encircled with a red line. Note that due to a lack of common definitions and concepts across disciplines, this figure is neither complete nor are the terms used unambiguous.}
	\label{img:Figure02}
\end{figure}

In order to assess a system from the perspective of IT security, the three main security goals\footnote{We note that the concepts covered by the terms availability and integrity differ to some extent from the ones they usually denote. Indeed, prevalent attacks on availability are the result of a large-scale violation of integrity of the system's output data. However, this usage has widely been adopted in the research area.} are used, which may all be targeted by attackers \cite{Papernot2016d, Biggio2018}:
\begin{enumerate}
\item Confidentiality, the protection of data against unauthorised access. A successful attack may for instance uncover training data in medical AI prognostics.
\item Availability, the guarantee that IT services or data can always be used as intended. A successful attack may for instance make AI-based spam filters block legitimate messages, thus hampering their normal operation.
\item Integrity, the guarantee that data are complete and correct and have not been tampered with. A successful attack may for instance make AI systems produce specific wrong outputs.
\end{enumerate}
This article focuses on integrity, cf. \autoref{img:Figure02}, since this is the most relevant threat in the computer vision applications cited above, which motivate our interest in the topic. Confidentiality and availability are thus largely out of scope. Nevertheless, further research in their direction is likewise required, since in other applications attacks on these security goals may also have far-reaching consequences, as can be seen by the short examples mentioned above.
Besides the three security goals, an AI system has to be assessed in terms of many additional aspects, cf. \autoref{img:Figure02}. While this paper is focused on the integrity of the AI model and the data used, it also touches important related aspects, such as robustness, interpretability and documentation.

\subsection{Related work}\label{sec:related}
Although the broader AI community remains largely unaware of the security issues involved in the use of AI systems, this topic has been studied by experts for many years now. Seminal works, motivated by real-world incidents, were concerned with attacks and defences for simple classifiers, notably for spam detection \cite{Dalvi2004, Lowd2005, Barreno2006, Biggio2013}. The field witnessed a sharp increase in popularity following the first publications on adversarial examples for deep neural networks (\cite{Szegedy2013, Goodfellow2014}, cf. \autoref{sec:operation}). Since then, adversarial examples and data poisoning attacks (where an attacker manipulates the training data, cf. \autoref{sec:poisoning}) have been the focus of numerous publications. Several survey articles \cite{Papernot2016d, Biggio2018, Liu2018b, Xu2019} provide a comprehensive overview of attacks and defences on the AI level.

Research on verifying and proving the correct operation of AI systems has also been done, although it is much scarcer \cite{Huang2017a, Katz2017, Gehr2018, Singh2019}. One approach to this problem is provided by the area of explainable AI (XAI, cf. \autoref{sec:ai_specific_mitigation}), which seeks to make decisions taken by an AI system comprehensible to humans and thus to mitigate an essential shortcoming of cAI systems.

Whereas previous survey articles like the ones cited above focus on attacks and immediate countermeasures on the level of the AI system itself, our publication takes into account the whole data supply chain (cf. \autoref{sec:supply}) and the fact that the AI system is just part of a larger IT system. On the one hand, for doing so, we draw up a more complete list of attacks which might ultimately affect the AI system. On the other hand, we argue that defences should not only be implemented in the AI systems themselves. Instead, more general technical and organisational measures must also be considered (as briefly noted in \cite{Gilmer2018}) and in particular new AI-specific defences have to be combined with classical IT security measures.

\subsection{Outline}
The outline of the paper is as follows: 
First, we inspect the supply chain of cAI systems in detail in \autoref{sec:supply}, identifying and analysing vulnerabilities. AI-specific vulnerabilities are further analysed in \autoref{sec:ai_issues} in order to give some intuition about their root causes which are not already familiar from other IT systems. Subsequently, \autoref{sec:mitigation} sets out to present mitigations to the threats identified in \autoref{sec:supply}, focusing not only on the level of the AI system itself but taking a comprehensive approach. We conclude in \autoref{sec:conclusion}, where we touch on future developments and the crucial aspect of verifying correct operation of an AI system.

%% file: 30-supply_chain.tex
\section{Generalised AI supply chain}\label{sec:supply}
\begin{figure}
	\centering
	\includegraphics[width=\textwidth]{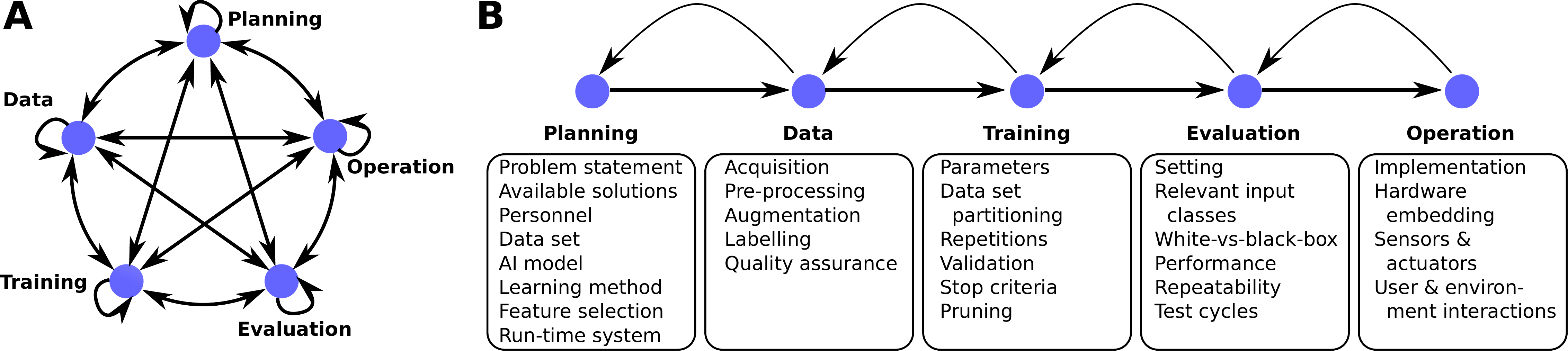}
	\caption{The development of cAI applications may be broken down into phases. (A) In reality, the development process is non-sequential, often relies on intuition and experience and involves many feedback loops on different levels. The developer tries to find the quickest route to an operational AI system with the desired properties. (B) For a simplified presentation, sequential phases are depicted. Here prominent functional components are shown for each phase. Besides this functional perspective, the phases may be considered in terms of robustness, data protection, user acceptance or other aspects.}
	\label{img:Figure03}
\end{figure}

In this section, we perform a detailed walk through the supply chain of cAI systems (cf. \autoref{img:Figure03}), mostly adopting the point of view of functionality or IT security. At each step of the supply chain, we identify important factors impacting the performance of the model and analyse possible vulnerabilities. Since our objective is to provide a comprehensive overview, we discuss both classical vulnerabilities well-known from traditional IT systems as well as qualitatively new attacks which are specific to AI systems. Whereas classical vulnerabilities should be addressed using existing evaluation and defence methods, AI-specific attacks additionally require novel countermeasures, which are discussed in this section to some extent, but mostly in \autoref{sec:mitigation}.

The supply chain we consider for our analysis is that of a generalised AI application. This approach is useful in order to get the whole picture at a suitable level of abstraction. We note, however, that concrete AI applications, in particular their boundary conditions, are too diverse to consider every detail in a generalised model. For instance, AI systems can be used for making predictions from structured and tabular data, for computer vision tasks and for speech recognition but also for automatic translation or for finding optimal strategies under a certain set of rules (e.g. chess, go). For anchoring the generalised analysis in concrete use cases, specific AI applications have to be considered. It may hence be necessary to adapt the general analysis to the concrete setting in question or at least to the broader application class it belongs to. In the following, we use the example of traffic sign recognition several times for illustrating our abstract analysis.

\subsection{Planning}\label{sec:planning}
The first step that is required in the development of an operational AI system is a thorough problem statement answering the question which task has to be solved under which boundary conditions. Initially, the expected inputs to the system as well as their distribution and specific corner cases are defined and the required performance of the system with respect to these inputs is estimated, including:
\begin{itemize}
\item the accuracy, or some other appropriate metric to assess the correctness of results of the system,
\item the robustness, e.g. with respect to inputs from a data distribution not seen during training, or against maliciously crafted inputs,
\item the restrictions on computing resources (e.g. the system should be able to run on a smartphone) and
\item the runtime, i.e. combined execution time and latency.
\end{itemize}
Next, it might be helpful to analyse if the problem at hand can be broken down into smaller sub-tasks which could each be solved on their own. One may hope that the resulting modules are less complex compared to a monolithic end-to-end system and, therefore, are better accessible for interpretation and monitoring. Once the problem and the operational boundary conditions have been clearly defined, the state of the art of available solutions to related problems is assessed. Subsequently, one or several model classes and ML algorithms (e.g. back-propagation of error \cite{Werbos19882}) for training the models are chosen which are assumed to be capable of solving the given task. In case a model class based on neural networks is chosen, a pre-trained network might be selected as a base model. Such a network has been trained beforehand on a possibly different task with a large data set (e.g. ImageNet \cite{imagenet}) and is used as a starting point in order to train the model for solving the task at hand using transfer learning. Such pre-trained networks (e.g. BERT \cite{Delvin2019} in the context of natural language processing) can pose a security threat to the AI system if they are modified or trained in a malicious way as described in sections \ref{sec:data_acquisition}, \ref{sec:VulnerabilitiesTraining}.

Based on the choices made before, the required resources in terms of quantity and quality (personnel, data set, computing resources, hardware, test facilities etc.) are defined. This includes resources required for threat mitigation (cf. \autoref{sec:mitigation}). Appropriate preparations for this purpose are put into effect. This applies in particular to the documentation and cryptographic protection of intermediate data, which affects all phases up until operation.

In order to implement the model and the ML algorithm, software frameworks (e.g. TensorFlow, PyTorch, sklearn \cite{tensorflow, pytorch, scikit}) might additionally be used in order to reduce the required implementation effort. This adds an additional risk in the form of possible bugs or backdoors which might be contained in the frameworks used.

\subsection{Data acquisition and preprocessing}\label{sec:data_acquisition}
After fixing the boundary conditions, appropriate data for training and testing the model need to be collected and preprocessed in a suitable way. To increase the size of the effective data set without increasing the resource demands, the data set may be augmented by both transformations of the data at hand and synthetic generation of suitable data. The acquisition can start from scratch or rely on an existing data set. In terms of efficiency and cost, the latter approach is likely to perform better. However, it also poses additional risks in terms of IT security, which need to be assessed and mitigated.

Several properties of the data can influence the performance of the model under normal and adverse circumstances. Using a sufficient quantity of data of good quality is key to ensuring the model's accuracy and its ability to generalise to inputs not seen during training. Important features related to the quality of data are, in a positive way, the correctness of their labels (in the setting of supervised learning) and, in a negative way, the existence of a bias. If the proportion of wrongly labelled data (also called noisy data) in the total data set is overly large, this can cripple the model's performance. If the training data contain a bias, i.e. they do not match the true data distribution, this adversely affects the performance of the model under normal circumstances. In special cases it might be necessary though to use a modified data distribution in the training data to adequately consider specific corner cases. Furthermore, one must ensure that the test set is independent from the training set in order to obtain reliable information on the model's performance. To trace back any problems that arise during training and operation, a sufficient documentation of the data acquisition and preprocessing phase is mandatory.

\subsubsection{Collecting data from scratch}
A developer choosing to build up his own data set has more control over the process, which can make attacks much more difficult. A fundamental question is whether the environment from which the data are acquired is itself controlled by the developer or not. For instance, if publicly available data are incorporated into the data set, the possibility of an attacker tampering with the data in a targeted way may be very small, but the extraction and transmission of the data must be protected using traditional measures of IT security. These should also be used to prevent subsequent manipulations in case an attacker gets access to the developer's environment. In addition, the data labelling process must be checked to avoid attacks. This includes a thorough analysis of automated labelling routines and the reliability of the employees that manually label the data as well as checking random samples of automatically or externally labelled data. Moreover, when building up the data set, care must be taken that it does not contain a bias.

\subsubsection{Using existing data}\label{sec:poisoning}
If an existing data set is to be used, the possibilities for attacks are diverse. If the developer chooses to acquire the data set from a trusted source, the integrity and authenticity of the data must be secured to prevent tampering during transmission. This can be done using cryptographic schemes. 

Even if the source is deemed trustworthy, it is impossible to be sure that the data set is actually correct and has not fallen prey to attacks beforehand. In addition, the data set may be biased, and a benign but prevalent issue may be data that were unintentionally assigned wrong labels (noise in the data set may be as high as 30\% \cite{Wang2018c,Veit2017}). The main problem in terms of IT security are so-called poisoning attacks though. In a poisoning attack, the attacker manipulates the training set in order to influence the model trained on this data set. Such attacks can be divided into two categories: 

\begin{figure}
	\centering
	\includegraphics[width=\textwidth]{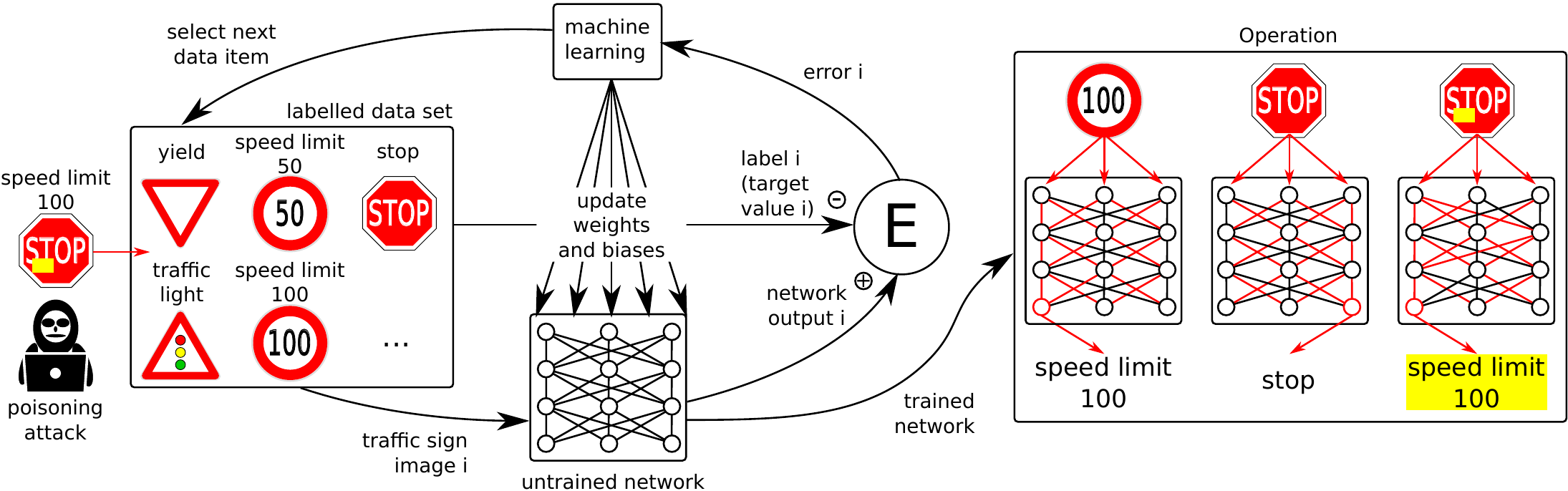}
	\caption{A so-called poisoning or backdooring attack may be mounted by an attacker if he gets the chance to inject one or more manipulated data items into the training set: the manipulated data lead to undesired results but the usual training and test data still produce the desired results, making it extremely hard to detect backdoors in neural networks. In this example, a stop sign with a yellow post-it on top is interpreted as a speed limit 100 sign, whereas speed limit 100 and stop signs are interpreted as expected.}
	\label{img:Figure04}
\end{figure}
\begin{enumerate}
 \item Attacks on availability: The attacker aims to maximise the generalisation error of the model \cite{Xiao2014}, \cite{Biggio2014b}, \cite{Mei2015} by poisoning the training set. This attack can be detected in the testing phase since it decreases the model's accuracy. A more focused attack might try to degrade the accuracy only on a subset of data. For instance, images of stop signs could be targeted in traffic sign recognition. Such an attack would only affect a small fraction of the test set and thus be more difficult to detect. The metrics used for testing should hence be selected with care.
 \item Attacks on integrity: The attacker aims to introduce a backdoor into the model without affecting its overall accuracy \cite{Chen:2017, Turner2019, Saha2019}, cf. \autoref{img:Figure04}, which makes it very hard to detect. The attack consists in injecting a special trigger pattern into the data and assigning it to a target output. A network trained on these data will produce the target output when processing data samples containing the trigger. Since the probability of natural data containing the trigger is very low, the attack does not alter the generalisation performance of the model. In classification tasks, the trigger is associated with a target class. For instance, in biometric authentication the trigger may consist in placing a special pair of sunglasses upon the eyes in images of faces. The model would then classify persons wearing these sunglasses as the target class.
\end{enumerate}

\subsection{Training}\label{sec:VulnerabilitiesTraining}
In this phase, the model is trained using the training data set and subject to the boundary conditions fixed before. To this end, several hyperparameters (number of repetitions, stop criteria, learning rate etc.) have to be set either automatically by the ML algorithm or manually by the developer and the data set has to be partitioned into training and test data in a suitable way. Attacks in this phase may be mounted by attackers getting access to the training procedure, especially if training is not done locally, but using an external source, e.g. in the cloud \cite{Gu2017}. Possible threats include augmenting the training data set with poisoned data to sabotage training, changing the hyperparameters of the training algorithm or directly changing the model's parameters (weights and biases). Furthermore, an attacker may manipulate already trained models. This can, for instance, be done by retraining the models with specially crafted data in order to insert backdoors, which does not require access to the original training data (trojaning attacks \cite{Ji2019}, \cite{Liu2018c}). A common feature of these attacks is that they assume a rather powerful attacker having full access to the developer's IT infrastructure. They can be mitigated using measures from traditional IT security for protecting the IT environment. Particular countermeasures include, on the one hand, integrity protection schemes for preventing unwarranted tampering with intermediate results as well as comprehensive logging and documentation of the training process. On the other hand, the reliability of staff must be checked to avoid direct attacks by or indirect attacks via the developers.

\subsection{Testing and evaluation}
After training, the performance of the model is tested using the validation data set and the metrics fixed in the planning phase. If it is below the desired level, training needs to be restarted and, if necessary, the boundary conditions need to be modified. This iterative process needs to be repeated until the desired level of performance is attained, cf. \autoref{img:Figure01} (B) and \autoref{img:Figure03} (A). In order to check the performance of the model, the process of evaluation needs to be repeated after every iteration of training, every time that the model goes into operation as part of a more complex IT system, and every time that side conditions change.

After finishing the training and validation phase, the test set is used for measuring the model's final performance. It is important that using the test set only yields heuristic guarantees on the generalisation performance of the model, but does not give any formal statements on the correctness or robustness of the model, nor does it allow understanding the decisions taken by the model if the structure of the model does not easily lend itself to human interpretation (black-box model). In particular, the model may perform well on the test set by having learnt only spurious correlations in the training data. Care must hence be taken when constructing the test set. A supplementary approach to pure performance testing is to use XAI methods (cf. \autoref{sec:ai_specific_mitigation}), which have often been used to expose problems which had gone unnoticed in extensive testing \cite{Lapuschkin2017}.

\subsection{Operation}\label{sec:operation}
A model that has successfully completed testing and evaluation may go into operation. Usually, the model is part of a more complex IT system, and mutual dependencies between the model and other components may exist. For instance, the model may be used in a car for recognising traffic signs. In this case, it receives input from sensors within the same IT system, and its output may in turn be used for controlling actuators. The embedded model is tested once before practical deployment or continuously via a monitoring process. If necessary, one can adjust its embedding or even start a new training process using modified boundary conditions and iterate this process until achieving the desired performance.

Classical attacks can target the system at different levels and impact the input or output of the AI model without affecting its internal operation. Attacks may be mounted on the hardware \cite{Clements2018} and operating system level or concern other software executed besides the model. Such attacks are not specific to AI models and are thus not in the focus of this publication. They need to be mitigated using classical countermeasures for achieving a sufficient degree of IT security. Due to the black-box property of AI systems, however, these attacks can be harder to detect than in a classical setting.

A qualitatively new type of attacks, called evasion attacks, focuses on AI systems, cf. \autoref{img:Figure05}. Evasion attacks have been well known in adversarial ML for years \cite{Biggio2018}. In the context of deep learning, these attacks are called adversarial attacks. Adversarial attacks target the inference phase of a trained model and perturb the input data in order to change the output of the model in a desired way \cite{Szegedy2013, Goodfellow2014}. Depending on the attacker's knowledge, adversarial attacks can be mounted in a white-box or grey-box setting:

\begin{figure}
	\centering
	\includegraphics[width=\textwidth]{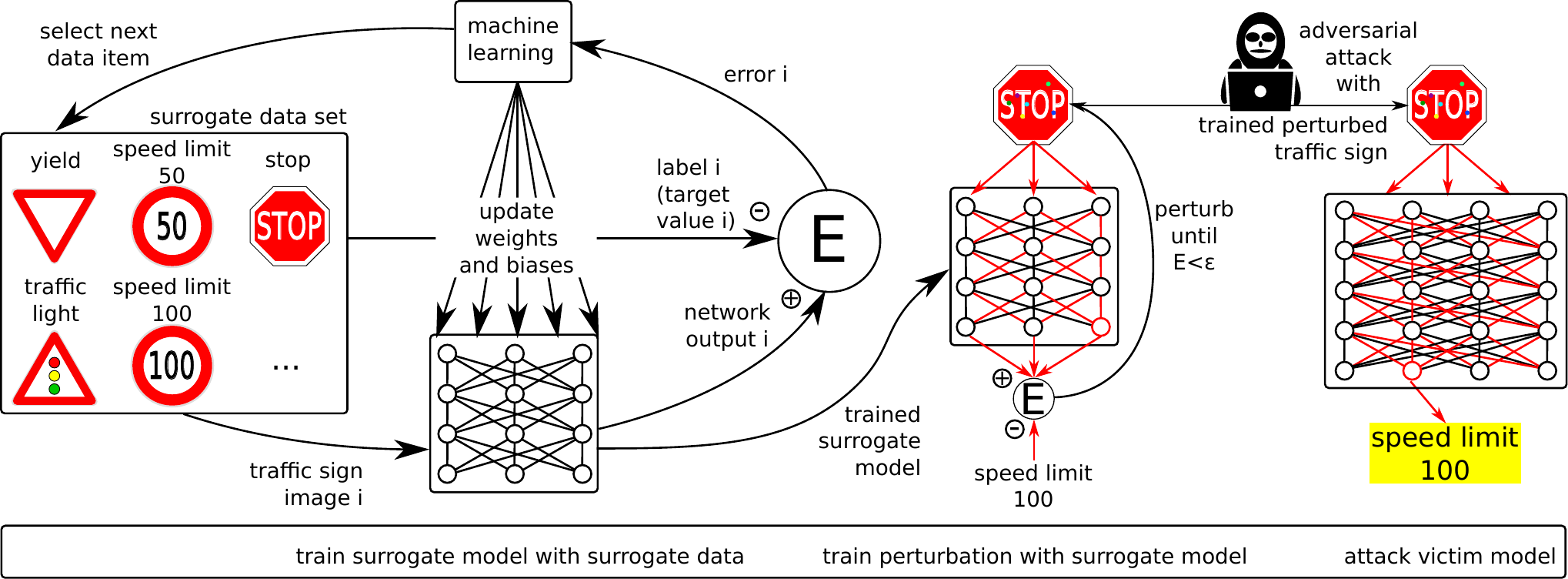}
	\caption{Adversarial attacks may be conducted without white-box access to the victim model: First, a surrogate model is trained using a surrogate data set. Labels for this data set might optionally be obtained via queries to the victim model. Subsequently, the trained surrogate model is used to generate adversarial input examples. In many cases, these adversarial examples may then be used successfully for attacking the victim model.}
	\label{img:Figure05}
\end{figure}
\begin{enumerate}
 \item In white-box attacks, the attacker has complete information about the system, including precise knowledge of defence mechanisms designed to thwart attacks. In most cases, the attacker computes the perturbation using the gradient of the targeted model. The Fast Gradient Sign Method of \cite{Goodfellow2014} is an early example, which was later enhanced by stronger attacks designed to create the perturbation in an iterative manner \cite{Carlini2017,Chen2017,Madry2018,Papernot2016c}.
 \item In grey-box attacks, the attacker does not have access to the internals of the model and might not even know the exact training set, although some general intuition about the design of the system and the type of training data needs to be present, as pointed out by \cite{Biggio2018}. In this case, the attacker trains a so-called surrogate model using data whose distribution is similar to the original training data and, if applicable, queries to the model under attack \cite{Papernot2016}. If the training was successful, the surrogate model approximates the victim model sufficiently well to proceed to the next step. The attacker then creates an attack based on the surrogate model, which is likely to still perform well when applied to the targeted model, even if the model classes differ. This property of adversarial examples, which is very beneficial for attackers, has been termed transferability \cite{Papernot2016f}.
\end{enumerate}

Adversarial attacks usually choose the resulting data points to be close to the original ones in some metric, e.g. the Euclidean distance. This can make them indistinguishable from the original data points for human perception and thus impossible to detect by a human observer. However, some researchers have raised the question whether this restriction is really necessary and have argued that in many applications it may not be \cite{Gilmer2018}. This applies in particular to applications where human inspection of data is highly unlikely and even blatant perturbations might well go unnoticed, as e.g. in the analysis of network traffic.

In most academic publications, creating and deploying adversarial attacks is a completely digital procedure. For situated systems acting in the sensory-motor loop, such as autonomous cars, this approach may serve as a starting point for investigating adversarial attacks but generally misses out on crucial aspects of physical instantiations of these attacks: First, it is impossible to foresee and correctly simulate all possible boundary conditions as e.g. viewing angles, sensor pollution and temperature. Second, sufficiently realistic simulations of the interaction effects between system modules and environment are hard to carry out. Third, this likewise applies to simulating individual characteristics of hardware components that influence the behaviour of these components. This means the required effort for generating physical adversarial attacks that perform well is much larger as compared to their digital copies. For this reason, such attacks are less well studied, but several publications have shown they can still work, in particular if attacks are optimised for high robustness to typically occurring transformations (e.g. rotation and translation in images) \cite{Athalye2018a,Brown2017,Evtimov2018,Eykholt2018b,Eykholt2017,Sharif2016}.

%% file: 40-vulnerabilities.tex
\section{Root causes of AI-specific vulnerabilities}\label{sec:ai_issues}
As described in \autoref{sec:supply}, AI systems can be attacked on different levels. Whereas many of the vulnerabilities are just variants of more general problems in IT security, which affect not only AI systems, but also other IT solutions, two types of attacks are specific to AI, i.e. poisoning attacks and adversarial examples (also known as evasion attacks). This section aims to give a general intuition of the fundamental properties specific to AI which enable and facilitate these attacks, and to outline some general strategies for coping with them.

\subsection{Huge input and state spaces}\label{sec:size_space}
Complex AI models contain many millions of parameters (weights and biases), which are updated during training in order to approximate a function for solving the problem at hand. As a result, the number of possible combinations of parameters is enormous and decision boundaries between input data where the models' outputs differ can only be approximate, cf. \autoref{tab:keyFactsCommonlyUsedArchitectures}. Besides, due to the models' non-linearity small perturbations in input values may result in huge differences in the output.
\begin{table}[]
\centering
\begin{tabular}{|l|l|l|l|l|l|}
\hline
model    & \parbox[t]{2cm}{number of\\distinct possible inputs} & \parbox[t]{2cm}{input size\\ (in bit)} & \parbox[t]{2cm}{output size\\ (in bit)}         & \parbox[t]{1.7cm}{number of\\ parameters} & \parbox[t]{1.7cm}{number of\\ layers} \\ \hline
LeNet-5 \cite{Lecun1998} & $2^{6272}$ &\parbox[t]{2cm}{$28 \cdot 28 \cdot 8\\ =6272$\\}& $10 \cdot 32$ & $\approx60\,$K   & 7       \\ \hline
VGG-16 \cite{Simonyan15} & $2^{1204224}$ & \parbox[t]{2cm}{$224 \cdot 224 \cdot 3 \cdot 8\\=1204224$\\}& $1000 \cdot 32$ & $\approx135\,$M   & 16       \\ \hline
ResNet-152 \cite{He2016} & $2^{1204224}$ & \parbox[t]{2cm}{$224 \cdot 224 \cdot 3 \cdot 8\\=1204224$\\}& $1000 \cdot 32$ &  $\approx60\,$M    & 152      \\ \hline
BERT \cite{Delvin2019} & $\leq 2^{7680}$ & \parbox[t]{2cm}{$\leq 512 \cdot 15\\=7680$\\}  & $\leq 512 \cdot 1000 \cdot 32$     & $\approx345\,$M   & 24       \\ \hline
\end{tabular}
\caption{The size of the input and state spaces of commonly used architectures in the field of object recognition (LeNet-5, VGG-16, ResNet-152) and natural language processing (BERT) is extremely large.}
\label{tab:keyFactsCommonlyUsedArchitectures}
\end{table}
In general, AI models are trained on the natural distribution of the data considered in the specific problem (e.g. the distribution of traffic sign images). This distribution, however, lies on a very low-dimensional manifold as compared to the complete input space (e.g. all possible images of the same resolution), which is sometimes referred to as the `curse of dimensionality'. \autoref{tab:keyFactsCommonlyUsedArchitectures} shows that the size of the input space for some common tasks is extremely large. Even rather simple and academic AI models as e.g. LeNet-5 for handwritten digit recognition have a huge input space. As a consequence, most possible inputs are never considered during training.

On the one hand, this creates a safety risk if the model is exposed to benign inputs which sufficiently differ from those seen during training, such that the model is unable to generalise to these new inputs. The probability of this happening depends on many factors, including the model, the algorithm used and especially the quality of the training data.

On the other hand, what is much more worrying, inputs which reliably cause malfunctioning for a model under attack, i.e. adversarial examples, can be computed efficiently and in a targeted way. Although much work has been invested in designing defences since adversarial examples first surfaced in deep learning, as of now, no general defence method is known which can reliably withstand adaptive attackers. That is, defences may work if information about their mode of operation is kept secret from an attacker. As soon as an attacker gains this information, which should in most cases be considered possible, he is able to overcome them. 

Besides the arms race in practical attacks and defences, adversarial attacks have also sparked interest from a theoretical perspective. Several publications deal with their essential characteristics. As pointed out by \cite{Biggio2018}, adversarial examples commonly lie in areas of negligible probability, blind spots where the model is unsure about its predictions. Furthermore, they arise by adding highly non-random noise to legitimate samples, thus violating the implicit assumption of statistical noise that is made during training. \cite{Khoury2018} relates adversarial examples to the high dimension of the input space and the curse of dimensionality, which allows constructing adversarial examples in many directions off the manifold of proper input data.  In \cite{Ilyas2019}, the existence of adversarial examples is ascribed to so-called non-robust features in the training data, which would also provide an explanation for their transferability property. By practical experiments \cite{Madry2018} demonstrate defences from the point of view of robust optimisation that show comparatively high robustness against strong adversarial attacks. Additionally and in contrast to most other publications, theses defenses provide some theoretical guarantee against a whole range of both static and adaptive attacks.

\begin{figure}
	\centering
	\includegraphics[width=1.0\textwidth]{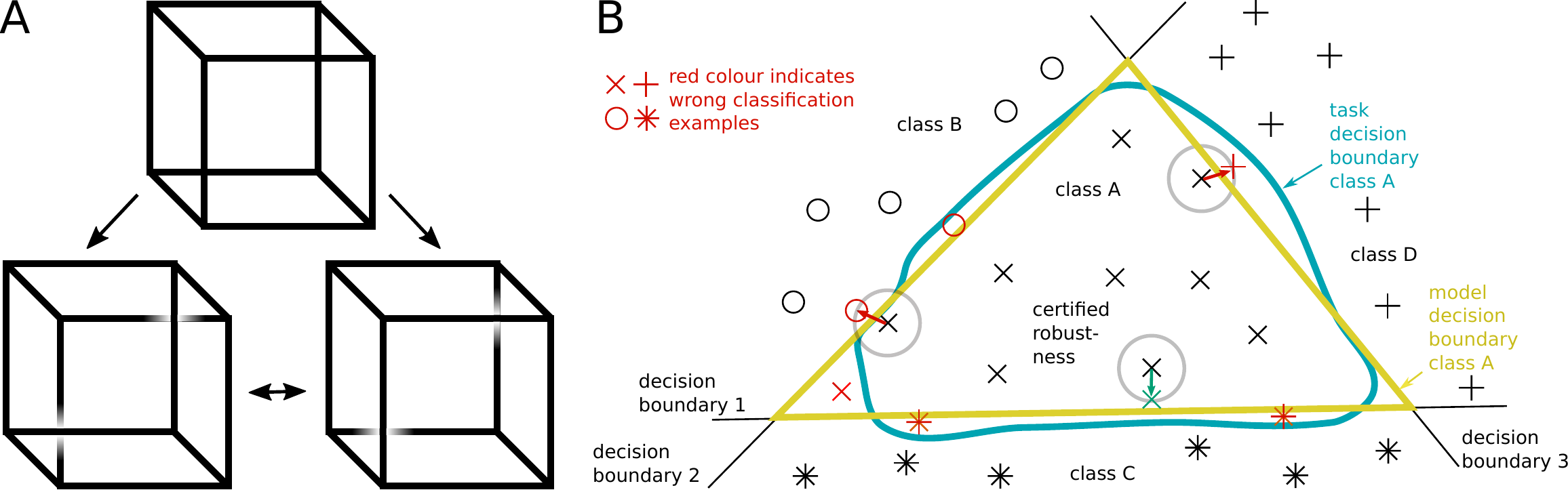}
	\caption{Error-prone decision-making in humans (A) and AI systems (B) as exemplified by the Necker cube as an example of an ambiguous image (A) and a schematic depiction of adversarial examples in a 2D-projection of state space (B). Task and model decision boundaries do not perfectly match and small changes in input may result in large changes in output. More details are given in the main text.}
	\label{img:Figure06}
\end{figure}

\autoref{img:Figure06} illustrates the problem of adversarial examples and its root cause and presents an analogy from human psychophysics. Decision-making in humans as well as in AI systems is error-prone since theoretically ideal boundaries for decision-making (task decision boundaries) are in practice instantiated by approximations (model decision boundaries). Models are trained using data (AI and humans) and evolutionary processes (humans). In the trained model, small changes in either sensory input or other boundary conditions (e.g. internal state) may lead to state changes whereby decision boundaries are crossed in state space, i.e. small changes in input (e.g. sensory noise) may lead to large output changes (here a different output class). Model and task decision, therefore, may not always match. Adversarial examples are found in those regions in input space where task and model decision boundaries differ, as depicted in \autoref{img:Figure06}:
\begin{itemize}
 \item Part A shows an example for human perception of ambiguous images, namely the so-called Necker cube: sensory input (image, viewpoint, lightening, ...), internal states (genetics, previous experience, alertness, mood, ...) and chance (e.g. sensory noise) determine in which of two possible ways the Necker cube is perceived: (top) either the square on the left/top side or the square on the right/bottom side is perceived as the front surface of the cube, and this perception may spontaneously switch from one to the other (bistability). Besides internal human states that influence which of the two perceptions is more likely to occur \cite{Ward2015}, the input image may be slightly manipulated such that either the left/top square (left) or the right/bottom square (right) is perceived as the front surface of the cube.
 \item Part B shows how all these effects are also observed in AI systems. This figure illustrates adversarial examples for a simplified two-dimensional projection of an input space with three decision boundaries forming the model decision boundary of class A (yellow) modelling the task decision boundary (blue): small modifications can shift (red arrows) input data from one model decision class to another, with (example on boundary 2 on the left) and without (example on boundary 3 on the right) changing the task decision class. Most data are far enough from the model decision boundaries to exhibit a certain amount of robustness (example on boundary 1 on the bottom). It is important to note that this illustration, depicting a two-dimensional projection of input space, does not reflect realistic systems with high-dimensional input space. In those systems, adversarial examples may almost always be found within a small distance from the point of departure. These adversarial examples rarely occur by pure chance but attackers may efficiently search for them.
\end{itemize}

\subsection{Black-box property and lack of interpretability}\label{sec:black_box}
A major drawback of complex AI models like deep neural networks is their shortcoming in terms of interpretability and explainability. Traditional computer programs solving a task are comprehensible and transparent at least to sufficiently knowledgeable programmers. Due to their huge parameter space as discussed in \autoref{sec:size_space}, complex AI systems do not possess this property. In their case, a programmer can still understand the boundary conditions and the approach to the problem; however, it is infeasible for a human to directly convert the internal representation of a deep neural network to terms allowing him to understand how it operates.
This is very dangerous from the perspective of IT security, since it means attacks can essentially only be detected from incorrect behaviour of the model (which may in itself be hard to notice), but not by inspecting the model itself. In particular, after training is completed, the model's lack of transparency makes it very hard to detect poisoning and backdooring attacks on the training data. For this reason, such attacks should be addressed and mitigated by thorough documentation of the training and evaluation process and by protecting the integrity of intermediate results or alternatively by using training and test data that have been certified by a trustworthy party.

A straightforward solution to the black-box property of complex AI models would be to use a model which is inherently easier to interpret for a human, e.g. a decision tree or a rule list \cite{Molnar2019}. When considering applications based on tabular data, for instance in health care or finance, one finds that decision trees or rule lists even perform better than complex cAI models in most cases \cite{Angelino2018, Rudin2018, Lundberg2020}, besides exhibiting superior interpretability. However, in applications from computer vision, which are the focus of this paper, or speech recognition, sAI models cannot compete with complex models like deep neural networks, which are unfortunately very hard to interpret. For these applications, there is hence a trade-off between model interpretability and performance. A general rule of thumb for tackling the issue of interpretability would still consist in using the least complex model which is capable of solving a given problem sufficiently well. Another approach for gaining more insight into the operation of a black-box model is to use XAI methods that essentially aim to provide their users with a human-interpretable version of the model's internal representation. This is an active field of research, where many methods have been proposed in recent years \cite{Samek2019, Molnar2019, Gilpin2018}. For more details, the reader is referred to \autoref{sec:ai_specific_mitigation}.


\subsection{Dependence of performance and security on training data}
The accuracy and robustness of an AI model is highly dependent on the quality and quantity of the training data. In particular, the model can only achieve high overall performance if the training data are unbiased. Despite their name, AI models currently used are not `intelligent', and hence they can only learn correlations from data but cannot by themselves differentiate spurious correlations from true causalities.

For economic reasons, it is quite common to outsource part of the supply chain of an AI model and obtain data and models for further training from sources which may not be trustworthy, cf. \autoref{img:Figure07}. On the one hand, for lack of computational resources and professional expertise, developers of AI systems often use pre-trained networks provided by large international companies or even perform the whole training process in an environment not under their control. On the other hand, due to the efforts required in terms of funds and personnel for collecting training data from scratch as well as due to local data protection laws (e.g. the GDPR in the European Union), they often obtain whole data sets in other countries. This does not only apply to data sets containing real data, but also to data which are synthetically created in order to save costs. Besides synthetic data created from scratch, this especially concerns data obtained by augmenting an original data set, e.g. using transformations under which the model's output should remain invariant.

Both these facts are problematic in terms of IT security, since they carry the risk of dealing with biased or poor-quality data and of falling prey to poisoning attacks (cf. \autoref{sec:supply}), which are very hard to detect afterwards. The safest way to avoid these issues is not to rely on data or models furnished by other parties. If this is infeasible, at least a thorough documentation and cryptographic mechanisms for protecting the integrity and authenticity of such data and models should be applied throughout the whole supply chain (cf. \autoref{sec:protect_evaluate}).

%% file: 50-mitigation.tex
\section{Mitigation of vulnerabilities of AI systems}\label{sec:mitigation}

\subsection{Assessment of attacks}
A necessary condition for properly reasoning about attacks is to classify them using high-level criteria. The result of this classification will facilitate a discussion about defences which are feasible and necessary. Such a classification is often referred to as a threat model or attacker model \cite{Papernot2016d, Biggio2018}.

An important criterion to consider is the \textbf{goal} of the attack. First, one needs to establish which security goal is affected. As already noted in \autoref{sec:introduction}, attackers can target either integrity (by having the system make wrong predictions on specific input data), availability (by hindering legitimate users from properly using the system) or confidentiality (by extracting information without proper authorisation). 
Besides, the scope of the attack may vary. An attacker may mount a targeted attack, which affects only certain data samples, or an indiscriminate one. In addition, the attacker may induce a specific or a general error. When considering AI classifiers, for instance, a specific error means that a sample is labelled as belonging to a target class of the attacker's choosing, whereas a general error only requires any incorrect label to be assigned to the sample.  
Furthermore, the ultimate objective of the attack must be considered. For example, this can be the unauthorised use of a passport (when attacking biometric authentication) or recognising a wrong traffic sign (in autonomous driving applications). In order to properly assess the attack, it is necessary to measure its real-world impact. For lack of more precise metrics commonly agreed upon, as a first step one might resort to a general scale assessing the attack as having low, medium or high impact.

The \textbf{knowledge} needed to carry out an attack is another criterion to consider. As described in \autoref{sec:VulnerabilitiesTraining}, an attacker has full knowledge of the model and the data sets in the white-box case. In this scenario, the attacker is strongest, and an analysis assuming white-box access thus gives a worst-case estimate for security. As noted in \cite{Carlini2019}, when performing such a white-box analysis, for the correct assessment of the vulnerabilities it is of paramount importance to use additional tests for checking whether the white-box attacks in question have been applied correctly, since mistakes in applying them have been observed many times and might yield wrong results.

In the case of a grey-box attack, conducting an analysis requires making precise assumptions on which information is assumed to be known to the attacker, and which is secret. \cite{Carlini2019} suggests that, in the same way as with cryptographic schemes, as little information as possible should be assumed to be secret when assessing the security of an AI system. For instance, the type of defence used in the system should be assumed to be known to the attacker. 


The third criterion to be taken into account is the \textbf{efficiency} of the attack, which influences the capabilities and resources an attacker requires. We assume the cost of a successful attack to be the most important proxy metric from the attacker's point of view. This helps in judging whether an attack is realistic in a real-world setting. If an attacker is able to achieve his objective using a completely different attack which does not directly target the AI system and costs less, it seems highly probable a reasonable attacker will prefer this alternative (cf. the concise discussion in \cite{Gilmer2018}). Possible alternatives may change over time though, and if effective defences against them are put into place, the attacker will update his calculation and may likely turn to attack forms he originally disregarded, e.g. attacks on the AI system as discussed in this paper.

The cost of a successful attack is influenced by several factors. First, the general effort and scope of a successful attack have a direct influence. For instance, the fact whether manipulating only a few samples is sufficient for mounting a successful poisoning attack or whether many samples need to be affected can have a strong impact on the required cost, especially when taking into account additional measures for avoiding detection. Second, the degree of automation of the attack determines how much manual work and manpower is required. Third, the fact whether an attack requires physical presence or can be performed remotely is likewise important. For instance, an attack which allows only a low degree of automation and requires physical presence is much more costly to mount and especially to scale. Fourth, attacking in a real-world setting adds further complexity and might hence be more expensive than an attack in a laboratory setting, where all the side conditions are under control.

A fourth important criterion is the \textbf{availability of mitigations}, which may significantly increase the attacker's cost. However, mitigations must in turn be judged by the effort they require for the defender, their efficiency and effectiveness. In particular, non-adaptive defence mechanisms may provide a false sense of security, since an attacker who gains sufficient knowledge can bypass them by modifying his attack appropriately. This is a serious problem pointed out in many publications, cf. \cite{Athalye2018, Gilmer2018}. As a rule, defence mechanisms should therefore respect Kerckhoffs' principle and must not rely on security by obscurity.

\subsection{General measures}\label{sec:protect_evaluate}

\begin{figure}
	\centering
	\includegraphics[width=\textwidth]{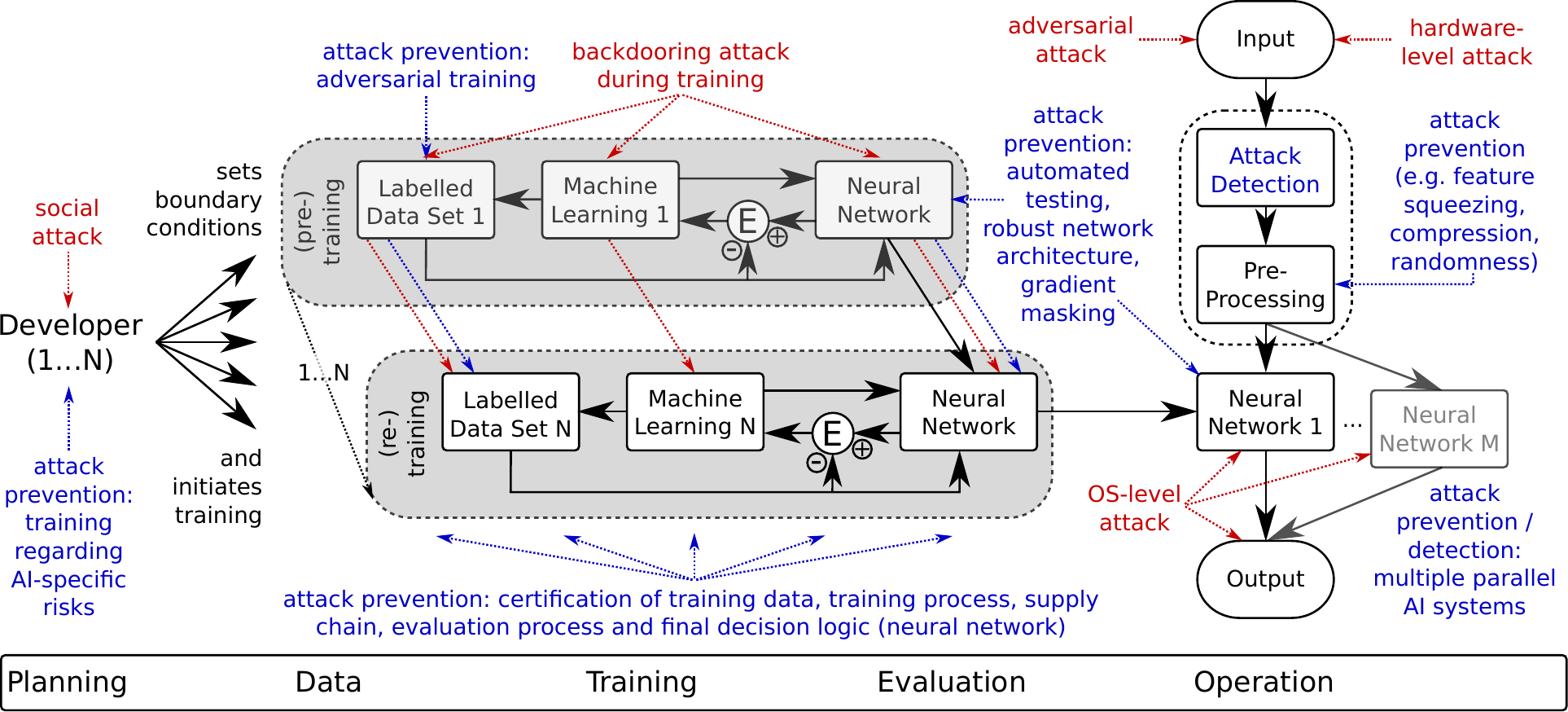}
	\caption{Summary of possible attacks (red) on AI systems and defences (blue) specific to AI systems depicted along the AI supply chain. Defences not specific to AI systems, e.~g. background checks of developers, hardware access control etc. are not shown here and should be adopted from classical IT security. Multiple AI training sessions with different data sets indicate the risk associated with pre-trained networks and externally acquired data.}
	\label{img:Figure07}
\end{figure}

A lot of research has been done on how to mitigate attacks on AI systems \cite{Carlini2019, robust_ml, robust_vision}. However, almost all the literature so far focuses on mitigations inside the AI systems, neglecting other possible defensive measures, and does not take into account the complete AI supply chain when assessing attacks. Furthermore, although certain defences like some variants of adversarial training \cite{Tramer2018, Salman2019} can increase robustness against special threat models, there is, as of now, no general defence mechanism which is applicable against all types of attacks. A significant problem of most published defences consists in their lack of resilience against adaptive attackers \cite{Athalye2018, Carlini2017a, Carlini2017b}. As already stated, the defence mechanisms used should be assumed to be public. The resistance of a defence against attackers who adapt to it is hence extremely important. In this section, we argue that a broader array of measures need to be combined for increasing security, especially if one intends to certify the safe and secure operation of an AI system, as seems necessary in high-risk applications like autonomous driving. An overview of defences and attacks is presented in \autoref{img:Figure07}.  

There is no compelling reason to focus solely on defending the AI system itself without taking into account additional measures which can hamper attacks by changing side conditions. This observation does not by any means imply that defences inside the AI system are unimportant or not necessary but instead emphasises that they constitute a last line of defence, which should be reinforced by other mechanisms.

\textbf{Legal measures} are most general. They cannot by themselves prevent attacks, but may serve as a deterrent to a certain extent, if properly implemented and enforced. Legal measures may include the adoption of new laws and regulation or specifying how existing laws apply to AI applications.

\textbf{Organisational measures} can influence the side conditions, making them less advantageous for an attacker. For instance, in biometric authentication systems at border control, a human monitoring several systems at once and checking for unusual behaviour or appearance may prevent attacks which can fool the AI system but are obvious to a human observer or can easily be detected by him if he is properly trained in advance. Restricting access to the development and training of AI systems for sensitive use cases to personnel which has undergone a background check is another example of an organisational measure. Yet another example is properly checking the identity of key holders when using a public key infrastructure (PKI) for protecting the authenticity of data.

\textbf{Technical measures outside the AI system} can be applied to increase IT security. The whole supply chain of collecting and preprocessing data, aggregating and transmitting data sets, pre-training models which are used as a basis for further training, and the training procedure itself can be documented and secured using classic cryptographic schemes like hash functions and digital signatures to ensure integrity and authenticity (this ultimately requires a PKI), preventing tampering in the process and allowing reproducing results and tracing back problems. Depending on the targeted level of security and traceability, the information covered may include all the training and test data, all AI models, all ML algorithms, a detailed logging of the development process (e.g. hyperparameters set by the developer, pseudo-random seeds, intermediate results) and comments of the developers concisely explaining and justifying each step in the development process. If the source of the data used is itself trusted, such documentation and cryptographic protection can later be validated to prove (with high probability) that no data poisoning attacks have been carried out, provided the validating party gets access to at least a sample of the original data and can check the correctness of intermediate results. As a further external technical measure, the AI system can be enhanced by using additional information from other sources. For example, in biometric authentication, biometric fakes can be detected using additional sensors \cite{Marcel2019}.

In a somewhat similar vein, the \textbf{redundant operation of multiple AI systems} running in parallel may serve to increase robustness to attacks, while at the same time increasing the robustness on benign data not seen during training.  These systems can be deployed in conjunction with each other and compare and verify each other's results, thus increasing redundancy. The final result might be derived by a simple majority vote, cf. \autoref{img:Figure07}. Other strategies are conceivable though. For instance, in safety-critical environments an alarm could be triggered in case the final decision is not unanimous and, if applicable, the system could be transferred to a safe fall-back state pending closer inspection. Increasing the redundancy of a technical system is a well-known approach for reducing the probability of undesired behaviour, whether due to benign reasons or induced by an attacker. However, the transferability property of adversarial examples (cf. \autoref{sec:operation}, \cite{Papernot2016f}) implies that attacks may continue to work even in the presence of redundancy, although their probability of success should at least slightly diminish. As a result, when using redundancy, one should aim to use conceptually different models and train them using different training sets that all stem from the data distribution representing the problem at hand, but have been sampled independently or at least exhibit only small intersections. While this does not in principle resolve the challenges posed by transferability, our intuition is that it should help to further decrease an attacker's probability of success.

\subsection{AI-specific measures}\label{sec:ai_specific_mitigation}
On the AI level, several measures can likewise be combined and used in conjunction with the general countermeasures presented above. First and foremost, appropriate state-of-the-art defences from the literature can be implemented according to their security benefits and the application scenario. One common approach for thwarting adversarial attacks is to make use of input compression \cite{Dziugaite2016,Das17}, which removes high-frequency components from input data that are typical for adversarial examples. More prominent still is a technique called \textbf{adversarial training}, which consists in precomputing adversarial examples using standard attack algorithms and incorporating them into the training process of the model, thus making it more robust and, in an ideal setting, immune to such attacks. State-of-the-art adversarial training methods may be identified using \cite{robust_ml, robust_vision, Madry2018}. In general, when dealing with countermeasures against adversarial attacks, it is important to keep in mind that many proposed defences have been broken in the past \cite{Athalye2018, Carlini2017a}, and that even the best defences available and combinations thereof \cite{Carlini2017b} may not fully mitigate the problem of adversarial attacks.

In terms of \textbf{defences against backdoor poisoning attacks} only a few promising proposals have been published in recent years \cite{Chen2018b, Tran2018, Wang2019}. Their main idea lies in the creation of a method which proposes possibly malicious data samples of the training set for manual examination. Those methods use the fact that a neural network trained on such a compromised data set learns the false classification of backdoored samples as exceptions, which can be detected from the internal representation of the network. It needs to be kept in mind though that those defences do not provide any formal guarantees and might be circumvented by an adaptive adversary.

As a first step, instead of preventing AI-specific attacks altogether, \textbf{reliably detecting} them might be a somewhat easier and hence more realistic task \cite{Carlini2017b}. In case an attack is detected, the system might yield a special output corresponding to this situation, trigger an alarm and forward the apparently malicious input to another IT system or a human in the loop for further inspection. It depends on the application in question whether this approach is feasible. For instance, asking a human for feedback is incompatible by definition with fully autonomous driving at SAE level 5 \cite{sae2018}.

\begin{figure}
	\centering
	\includegraphics[width=\textwidth]{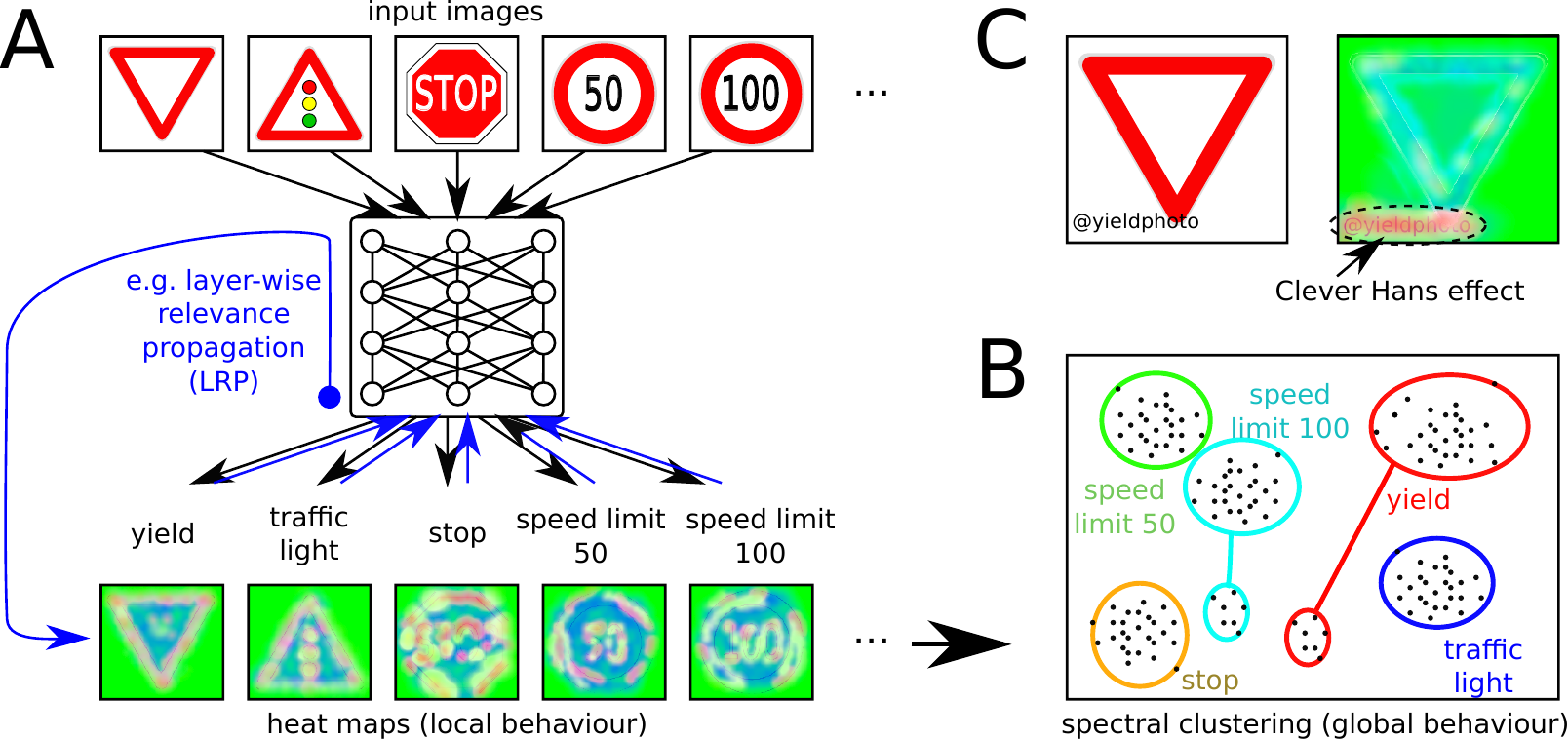}
	\caption{Schematic illustration of the application of explainable AI (XAI) methods to deduce (A) local and (B) global model behaviour of an AI system. (A) shows how heat maps are generated after labels were obtained for a specific input image, in this case using LRP \cite{Samek2016}, which assigns each input pixel a relative contribution to the output decision (green colours indicate lowest relevance, red colours highest relevance). (B) illustrates how many local model behaviour explanations are combined to explain global model behaviour, in this case using spectral analysis (cf. \cite{Lapuschkin2017}). Here multiple topographically distinct clusters for individual labels shown in a 2D projection of input space indicate some kind of problem: the small cluster for \textit{speed limit 100} represents the backdooring attack using modified stop signs (cf. \autoref{img:Figure04}) and the small cluster for the yield sign represents the Clever Hans effect illustrated in detail in (C), where specific image tags (here ``@yieldphoto'') correlate with specific input classes (here the yield sign) and the AI system focuses on these spurious correlations instead of causal correlations. Upon swapping the input data set (not containing any more spurious correlations of this kind), the AI model might show erroneous behaviour.}
	\label{img:Figure08}
\end{figure}

A different approach lies in using methods from the area of \textbf{explainable AI (XAI)} to better understand the underlying reasons for the decisions which an AI system takes, cf. \autoref{img:Figure08}. At the least, such methods may help to detect potential vulnerabilities and to develop more targeted defences. One example is provided by \cite{Lapuschkin2017}, which suggests a more diligent preprocessing of data for preventing the AI system from learning spurious correlations, which can easily be attacked. In principle, one can also hope that XAI methods will allow reasoning about the correctness of AI decisions under a certain range of circumstances. The field of XAI as focused on (deep) neural networks is quite young, and research has only started around 2015, although the general question of explaining decisions of AI systems dates back about 50 years \cite[pp.~41--49]{Samek2019}. So far, it seems doubtful there will be a single method which will fit in every case. Rather, different conditions will require different approaches. On the one hand, the high-level use case has a strong impact on the applicable methods: When making predictions from structured data, probabilistic methods are considered promising \cite{Molnar2019}, whereas applications from computer vision rely on more advanced methods like layer-wise relevance propagation (LRP) \cite{Bach2015, Montavon2015, Samek2016, Lapuschkin2017}. On the other hand, some methods provide global explanations, while others explain individual (local) decisions. It should be noted that by using principles similar to adversarial examples, current XAI methods can themselves be efficiently attacked \cite{Dombrowski2019}. Malicious input data leave the model output unchanged while completely altering the explanations provided. Based on theoretical observations, \cite{Dombrowski2019} suggests countermeasures for thwarting such attacks.

A third line of research linked to both other approaches is concerned with \textbf{verifying and proving} the safety and security of AI systems. Owing to the much greater complexity of this problem, results in this area, especially practically usable ones, are scarce \cite{Huang2017a, Katz2017,Gehr2018, Singh2019}. A general idea for harnessing the potential of XAI and verification methods may be applied, provided one manages to make these methods work on moderately small models. In this case, it might be possible to \textbf{modularise} the AI system in question so that core functions are mapped to small AI models \cite{Mascharka2018}, which can then be checked and verified. From the perspective of data protection, this approach has the additional advantage that the use of specific data may be restricted to the training of specific modules. In contrast to monolithic models, this allows unlearning specific data by replacing the corresponding modules \cite{Bourtoule2019}.

%% file: 60-conclusion.tex
\section{Conclusion and outlook}\label{sec:conclusion}
The supply chain of AI systems can give rise to malfunctions and is susceptible to targeted attacks at different levels. When facing naturally occurring circumstances and benign failures, i.e. in terms of safety, well-trained AI systems display robust performance in many cases. In practice, they may still show highly undesired behaviour, as exemplified by several incidents involving Tesla cars \cite{wiki_tesla}. The main problem in this respect is insufficient training data. The black-box property of the systems aggravates this issue, in particular when it comes to gaining user trust or establishing guarantees on correct behaviour of the system under a range of circumstances.

The situation is much more problematic though when it comes to the robustness to attacks exhibited by the systems. Whereas a lot of attacks can be combated using traditional measures of IT security, the AI-specific vulnerabilities to poisoning and evasion attacks can have grave consequences and do not yet admit reliable mitigations. Considerable effort has been put into researching AI-specific vulnerabilities, yet more is needed, since defences still need to become more resilient to attackers if they are to be used in safety-critical applications. In order to achieve this goal, it seems furthermore indispensable to combine defence measures at different levels and not only focus on the internals of the AI system.

Additional open questions concern the area of XAI, which is quite recent with respect to complex AI systems. The capabilities and limitations of existing methods need to be better understood, and reliable and sensible benchmarks need to be constructed to compare them \cite{Osman2020}. The topic of formal verification of the functionality of an AI system is an important enhancement that should further be studied. A general approach for obtaining better results from XAI and verification methods is to reduce complexity in the models to be analysed. We argue that for safety-critical applications the size of AI systems used for certain tasks should be minimised subject to the desired performance. If possible, one might also envision using a modular system containing small modules, which lend themselves more easily to analysis. A thorough evaluation using suitable metrics should be considered a prerequisite for the deployment of any IT system and, therefore, of any AI system.

Thinking ahead, the issue of AI systems which are continuously being trained using fresh data (called continual learning, \cite{Parisi2019}) also needs to be considered. This approach poses at least two difficulties as compared to the more static supply chain considered in this article. On the one hand, depending on how the training is done, an attacker might have a much better opportunity for poisoning training data. On the other hand, results on robustness, resilience to attacks or correctness guarantees will only be valid for a certain version of a model and may quickly become obsolete. This might be tackled by using regular checkpoints and repeating the countermeasures and evaluations, at potentially high costs.

Considering the current state of the art in the field of XAI and verification, it is unclear whether it will ever be possible to formally certify the correct operation of an arbitrary AI system and construct a system which is immune to the AI-specific attacks presented in this article. It is conceivable that both certification results and defences will continue to only yield probabilistic guarantees on the overall robustness and correct operation of the system. If this assumption turns out true for the foreseeable future, its implications for safety-critical applications of AI systems need to be carefully considered and discussed without bias. For instance, it is important to discuss which level of residual risk, if any, one might be willing to accept in return for possible benefits of AI over traditional solutions, and in what way the conformance to a risk level might be tested and confirmed. For instance, humans are required to pass a driving test before obtaining their driver's licence and being allowed to drive on their own. While a human having passed a driving test is not guaranteed to always respect the traffic rules, to behave correctly and to not cause any harm to other traffic participants, the test enforces a certain standard. In a similar vein, one might imagine a special test to be passed by an AI system for obtaining regulatory approval. In these cases the risks and benefits of using an AI system and the boundary conditions for which the risk assessment is valid should be made transparent to the user. However, the use of any IT system that cannot be guaranteed to achieve the acceptable risk level as outlined above could in extreme cases be banned for particularly safety-critical applications. Specifically, such a ban could apply to pure AI systems, if they fail to achieve such guarantees.


%
%

\section*{Acknowledgements}
We would like to thank Ute Gebhardt, Rainer Plaga, Markus Ullmann and Wojciech Samek for carefully proofreading earlier versions of this document and providing valuable suggestions for improvement. Further we would like to thank Frank Pasemann, Petar Tsankov, Vasilios Danos and the VdTÜV-BSI AI work group for fruitful discussions.

%% file: 70-literature.tex

\footnotesize
\bibliographystyle{plain}
\bibliography{ms}
